\def\beg{\begin{equation}}
\def\eeq{\end{equation}}
\documentstyle[12pt]{article}
\begin{document}
\begin{center}
{\Large{\bf Comments on ``Coexistence of Composite-Bosons 
and Composite-Fermions in Quantum Hall Bilayers, Simon et al 
cond-mat/0301203" }}
\vskip0.35cm
{\bf Keshav N. Shrivastava}
\vskip0.25cm
{\it School of Physics, University of Hyderabad,\\
Hyderabad  500046, India}
\end{center}

Simon et al suggest that composite fermions (CF) must be replaced by composite bosons (CB) and ``111" state is a boson state. However, we find that such a transition in real GaAs is not possible and the CF state can not become a boson. Simon et al suggest that the total 
number of CFs and CBs is conserved. However, we find that the number 
of quasiparticles need not be conserved. Similarly, Simon et al 
suggest that CFs are formed but we find that this formation does not conserve energy so that CFs will not be formed. Simon et al compute using products of fermion and boson wave functions. We find that 
these results are not in agreement with the experimental data.

\vfill
Corresponding author: keshav@mailaps.org\\
Fax: +91-402-301 0145.Phone: 301 0811.
\newpage
\baselineskip22pt
\noindent {\bf 1.~ Introduction}

     Recently, Simon et al[1] have  suggested that in a bilayer with each layer half filled as the layer separation is reduced, the CF Fermi sea must be replaced by a composite boson or a ``111" state. It has been proposed that CFs and CBs coexist in two interpenetrating fluids. It is further claimed that a Chern-Simons transport theory is constructed  that is compatible with experiments. The numerically exact computation of the energy 
and wave function has been presented. We find that (a) ``111" 
state is antisymmetric and hence a fermion. In a Laughlin 
representation both the fermions as well as bosons are possible. 
(b) The formation of CF and hence that of CB is 
subject to flux attachment to the electrons but there is no 
provision to detach flux from the electrons. We can make 
Laughlin's incompressible fractionally charged quasiparticles or 
the compressible CFs with field corrected but not both. The 
flux is not independent of currents. (c) We discuss the Chern-Simons field. However, we find that such a theory is not in agreement with 
the experimental data. The numerically exact calculation gives a very broad maxima whereas the experimental data has a very sharp peak.

\noindent{\bf 2.~~Comments.}

{\it (i) The ``111" state }

  Consider a system with two layers in which each layer is half filled, i.e., $\nu=n\phi_o/B$=1/2 where $n$ is the density of electrons per unit area and $\phi_o$ is the unit flux, $\phi_o$=hc/e. The magnetic field, $B$, is applied perpendicular to the surface of the sample. The spacing between two layers is $d$. Simon et al[1] suggest that for large $d$ the system is described as compressible composite fermion (CF) Fermi sea with strong intralayer correlations and no interlayer correlations. For small values of $d$, there is a ``111" state which can be described as a composite boson (CB). 

First of all the CFs are unrealistic and internally inconsistent objects as explained in the past[2]. Their mass is much too big and their size is also too large to fit in. The density of CFs also can not be equal to that of the electrons. The ``111" state will have factors like,
\beg
\psi_{111} \sim (z_{11}-z_{12})(z_{21}-z_{22})(z_{31}-z_{32})
\eeq
where in $z_{ij}$ the first subscript indicates coordinates of three electrons,$z_1$, $z_2$, and $z_3$. The second subscript identifies two electrons, $z_{i1}$ and $z_{i2}$. When the second subscript is interchanged $z_{11}$-$z_{12}$ becomes $z_{12}$-$z_{11}$ so that it changes sign. All of the three products change sign upon interchanging the second subscript. Hence $z_{11}$-$z_{12}$ is antisymmetric. Similarly, the other two products are also antisymmetric. The exponent $m$=1 or odd in (z$_{11}$-z$_{12})^m$ maintains antisymmetry. Therefore,
$``111"$ is antisymmetric and hence always a fermionic state. This state can not become a boson. When $m=3$ the state $``333"$ will also be antisymmetric and hence a fermionic state and never a bosonic state. When $m$=even, for example, $``222"$,i.e.,
\beg
\psi_{222}= (z_{11}-z_{12})^2(z_{21}-z_{22})^2(z_{31}-z_{32})^2
\eeq
is always a bosonic state by symmetry alone. The thermal distribution of bosons, $n_b$ and that of fermions, $n_f$, are given by,
\beg
<n_b> = [exp(\hbar\omega - \mu)/k_BT-1]^{-1}
\eeq
and
\beg
<n_f> = [exp(\epsilon-\epsilon_F)/k_BT+1]^{-1}
\eeq
at T=0, $\epsilon$-$\epsilon_F/k_B$T$\to\infty$ and hence the additional term of +1 in the denominator becomes negligible. Therefore, as far as thermal value is concerned,
\beg
<n_b>=<n_f> \,\,\,\,with\,\,\,\, \hbar\omega= \epsilon-\epsilon_F
\eeq
and the concept of the three dimensional fermi surface given up, in one-dimension, the $n_b$ and $n_f$ can cross over. In this naive interpretation, in one dimension at zero temperature the boson number density becomes equal to the fermi value. Otherwise the Fermi and Bose
statistices never meet. Therefore there is no chance  of transmutation of Fermi and Bose statistices. There are quantum mechanical reasons why Bose and Fermi statistices never cross. For example, the fermions obey the Pauli exclusion principle whereas bosons do not. In fact, Tomonaga[3] had realised that one can not assign numerical values simultaneously to both kind of occupation numbers. Therefore, $``111"$
will be a fermion and never become a boson. Simon et al consider the nature of phase transition between CF and CB. In the modern field theory this type of change of statistics is not permitted. The assertion of Simon et al[1] that $`111"$ state is a bosonic state is not correct.

{\it(ii) Number conservation.}

     Simon et al suggested that transition between the CF Fermi sea and the CB $``111"$ may be of first order. They also suggest that the total number of CBs and the CFs remain fixed. Usually the fermion number is conserved but the boson number need not be conserved. However, if bosons are charged with a unit charge, then charge conservation includes the conservation of boson number also. When one boson of fractional charge combines with one fermion of fractional charge, two particles combine into one, then charge is conserved but the number of particles is not. For example, charges 2/3 and 1/3 combine to make one particle of charge 1, then the number of quasiparticles is not conserved. Two particles of
charge 1/3 each may combine to form one particle of charge 2/3. Then the number is not conserved. If one electron and one hole combine and one photon is emitted, then the number of quasiparticles is not conserved.
Therefore, the experimental observation need not agree with the theoretical idea of Simon et al that the number of quasiparticles is conserved.

{\it(iii) Energy conservation.}

     The CF is obtained by attaching 2 zeros of the wave function to each electron[4-6]. This attachment has not been obtained by quantum mechanics. This is a result of $``random"$ writing of experimentally obtained effective charges which has not yet been derived by theory of electrodynamics. The Chern-Simons correction has been obtained by setting the scalar potential to zero and correcting only the vector potential of the electromagnetic field. This kind of calculation will not be consistent with the electromagnetic field theory of the classical electrodynamics which is the basis of production of magnets. It will modify the Maxwell equations and hence will be inconsistent with the experimental data. Simon et al suggest that the two flux quanta are attached to one electron so that the magnetic field becomes,
\beg
B^*=B-2n\phi_o
\eeq
and the electric field becomes,
\beg
E^*=E-2\epsilon J
\eeq
where ${\bf E}$ is the actual electric field, ${\bf J}$ is the fermion current and $\epsilon=(h/e^2)\tau$ with $\tau=i\sigma_y$. Here $E^*$=$\rho_f{\bf J}$ so that the electrical resistivity becomes,
\beg
\rho=\rho_f+2\epsilon.
\eeq
When $E^*\times B^*$ is the electromagnetic energy, it is less than $E\times B$. The energy in $E^*\times B^*$ is less than in $E\times B$
by the amount, $4\epsilon n \phi_o J_x$-2$\epsilon J_x B_z$-2$n\phi_oE_x$. Then what happened to the $``missing\,\,\, energy"$? This means that when CFs are created from electrons, the energy is not conserved. There is nothing in the Jain's formulas[3] which can conserve energy. Therefore, the CFs are not consistent with the $``principle\,\,\, of \,\,\,conservation \,\,\,of \,\,\,energy"$. When CF breaks, the electrons and even number of flux quanta are created. The flux quanta are bosons and their emission satisfies the Goldstone theorem but along with the flux quanta, currents should also be emitted. No such currents are prescribed by CF formulas. Hence, the CF is internally inconsistent.

{\it(iv) Incompressibility.}

     There is an idea that electron is bound to a single zero of the wave function. So that when product wave function is written, the electron wave function will be multiplied by zero and become zero.
Then there is no flux attachment in the Laughlin wave function. In the case of Laughlin's wave function, the system is incompressible. In the case of flux attachment, the system becomes compressible. Either the charge is modified by making the system incompressible or the field is modified in the compressible system. Thus both possibilities will be incompatible but one may resort to taking linear combination of both results but then such a phenomenon does not occur.

     Simon et al suggest that a Chern-Simons theory can be written for $``111"$ state. Here each electron is exactly modeled as a boson bound to 1  flux quantum, where the bosons see flux from both layers. The effective magnetic field seen due to electron density $n^{(1)}$ in one layer and $n^{(2)}$ in another layer is written as,
\beg
B^{*\alpha}= B - \phi_o(n^{(1)}+n^{(2)})
\eeq
and the electric field becomes,
\beg
E^{*\alpha}= E^\alpha-\epsilon(J^{(1)}+J^{(2)}).
\eeq
At the condensation point, $E^{*\alpha}=0$ so that $E^\alpha $=$\epsilon(J^{(1)}+J^{(2)})$. However, as pointed out, if both $E^*$ and $B^*$ are reduced, then energy is not conserved.

{\it(v) Calculation and Data.}

     Transition wavefunctions/numerical calculations. Simon et al[1] considered a finite sized bilayer sphere with 5 electrons per layer and a monopole of flux 9$\phi_o$ at its centre. They write products of wave functions in which some factors have even $m$ and some have odd so that products of fermion and boson wave functions arise. An effort is made to plot,
\beg
|<\psi_{trial}|\psi_{ground}>|^2 \,\,\, and \,\,\,E_{trial}-E_{ground}
\eeq
These theoretical results are not in accord with the experimental data of Spielman et al[7]. It has been suggested that the mixed CB-CF states have very high overlap with the exact ground state implying that the trial state is the ground state. So the trial state is a mixture of bosons and fermions. The experiment is consistent with the emission of a Goldstone boson which is not the same system as product of fermions
 and bosons. On theoretical grounds, there is no mention  of orthogonality
of wave functions and the orthogonal wave functions have not been written down. It has been suggested that CB-CF theory can be
 generalized for unequal densities as well as for filling fractions
 away from $\nu_1=\nu_2=1/2$. Surely, there are unequal densities
 but then such corrections if made will reduce the existing agreement between theory and the data. Simon et al have described the crossover between a CF liquid and a CB state and obtained exact results. They 
also claim that their appears to be a reasonable agreement with the data. The fermions obey anticommutators and bosons commutators but 
there is no way for anticommutators to become commutators. If 
fermions are transmutated to bosons, it may be so at zero 
temperature. The exactness of the Laughlin calculation assumes $``incompressibility"$ but the real system is compressible. In the experiments on quantum Hall effect in bilayers, there is no evidence
 of transmutation. The data need not be incompressible. There is no evidence of flux quanta attachment since there is no detachment. The experimental data also does not show evidence of conversion of bosons into fermions. In short, contrary to the claims made by Simon et al 
the data is not in agreemwent with the theory of Simon et al[1].

{\it(vi) Correction.}

    {\it The mass of a CF is several thousand times larger than the experimental mass. For $^2CF$ the mass may be 1000$m_e$ but the experimental mass is 0.4$m_e$. Hence there is no way for the CF-CB
theory to agree with the data.}

\noindent{\bf3.~~ Conclusions}.

     There is a product of fermion and boson wave functions and it is
claimed by Simon et al that such a wave function agrees with the data.
We find that the algebra in the Laughlin wave function assumes incompressibility while in the CF model the system is compressible.
The experimental data of Spielman et al has a sharp peak while the
calculation  of Simon et al has a broad response. Therefore, contrary to what is claimed the data are not in agreement with calculations of Simon et al.

\noindent{\bf About the author}: {\it Keshav Shrivastava has obtained Ph.D. degree from the
Indian Institute of Technology and D. Sc. from Calcutta University. 
He is a member of the American Physical Society, Fellow of the Institute of Physics (U.K.)and Fellow of the National Academy of Sciences, India. He has worked in the Harvard University, University of California at Santa Barbara, the University of Houston and the Royal Institute of Technology Stockholm. He has published 170 papers in the last 40 years. He is the author of two books}.

The correct theory of quantum Hall effect is given in ref.8.
\vskip1.25cm

\noindent{\bf5.~~References}
\begin{enumerate}
\item S. H. Simon, E. H. Rezayi and M. Milovanovic, cond-mat/0301203.
\item K. N. Shrivastava, cond-mat/0210320.
\item S. Tomonaga, Prog. Theoret. Phys. (Kyoto) {\bf 5}, 544 (1950).
\item J. K. Jain, Phys. Rev. Lett. {\bf 63}, 199 (1989).
\item K. Park and J. K. Jain, {\bf 80}, 437 (1998)
\item O. Heinonen, ed., Composite Fermions, World Scientific, 1998.
\item I. B. Spielman, J. P. Eisenstein, L. N. Pfeiffer and K. W. West, {\bf 84}, 5808(2000); {\bf87}, 036803(2001).
\item K.N. Shrivastava, Introduction to quantum Hall effect,\\ 
      Nova Science Pub. Inc., N. Y. (2002).
\end{enumerate}
\vskip0.1cm
Note: Ref. 8 is available from:\\
 Nova Science Publishers, Inc.,\\
400 Oser Avenue, Suite 1600,\\
 Hauppauge, N. Y.. 11788-3619,\\
Tel.(631)-231-7269, Fax: (631)-231-8175,\\
 ISBN 1-59033-419-1 US$\$69$.\\
E-mail: novascience@Earthlink.net

\vskip5.5cm
\end{document}